\documentclass[aps,twocolumn,preprintnumbers]{revtex4}
\usepackage{graphicx}
\newcommand{\EAS}{\hbox{EAS--1000}}
\begin{document}
\preprint{\texttt{astro-ph/0201343}}

\title{%
     Distribution of EAS arrival times
     according to data of the \EAS\
     prototype array%
\footnote{
     Proc.~27th International Conference on Cosmic Rays,
     Hamburg, 2001, \textbf{1}, 195--196.}
}

\author{
     Yu.\ A. Fomin, N. N. Kalmykov, G. V. Kulikov, V. I. Solovieva,
     V. P. Sulakov, O. V. Vedeneev, and M. Yu.\ Zotov
}

\affiliation{
     D. V. Skobeltsyn Institute of Nuclear Physics\\
     M. V. Lomonosov Moscow State University\\
     Moscow 119992, Russia
}
%\correspondence{N. N. Kalmykov\\ (kalm@eas10.eas.npi.msu.ru)}
%----------------------------------------------------------------------

\begin{abstract}
     We have analysed arrival times of extensive air showers (EAS)
     registered with the \EAS\ prototype array during the period from
     August 1997 till February 1999.
     Our analysis
%    that was based on certain non-parametric statistical methods
     has revealed that though the vast majority of samples of
     consecutive time intervals between EAS arrival
     times obey the exponential distribution, there are sequences of showers
     that have another distribution and thus violate the
     homogeneity hypothesis.
     The search for correlation between such events and clusters of
     showers and events with big delays between arrival times was also
     carried out.
\end{abstract}
\maketitle
%----------------------------------------------------------------------

\section{Introduction}

     Distribution of arrival times of extensive air showers (EAS) has
     been studied by methods of classical statistics by different
     research groups (see, e.g., Chikawa et~al.\ (1991) and Tsuji
     et~al.\ (1993)).  In this paper, we present some of the results
     of a similar investigation performed with the experimental data
     obtained with the \EAS\ prototype array. This installation
     operates at the Institute of Nuclear Physics of Moscow State
     University (Fomin et~al., 1999).

     The \EAS\ prototype array consists of eight detector units (DU)
     placed in the central part of the EAS MSU array and covers the
     area of $67\times22$~m.  The array is designed for registration
     of extensive air showers produced by cosmic rays with energy
     more than $10^{14}$~eV.

     When an EAS arrives, the corresponding signals from DU are
     transmitted to a PC for the preliminary program analysis of
     information quality and for EAS event verification.  The shower
     selection criterion is the triggering of any four adjacent
     detectors in the course of time gate less than 3.2~ns.
%----------------------------------------------------------------------

\section{Analysis of experimental data}

     203 days of twenty-four hour regular operation during the period
     from August 1997 to February 1999 were selected for the
     analysis.  The total number of showers in this data set is about
     $1.7\times10^6$.  The average time interval between successive
     showers is 10.5~s.  The discreteness of EAS arrival times is
     about 0.055~s (one tic of computer clock).

     The first step of verification of the hypothesis that a certain
     data sample obeys some definite distribution law is a test of
     homogeneity of the given sample (see, e.g., Bendat and Piersol
     (1986)).  There are numerous non-parametric criteria of
     homogeneity hypothesis.  To begin, we have chosen the series
     criterion, which is based on the analysis of the number of
     groups (series) of successive time intervals in the sample under
     consideration that have lengths greater than the median value
     and less than median value (Bol'shev and Smirnov, 1983; Bendat
     and Piersol, 1986).

     Using this criterion, we have verified a hypothesis about
     homogeneity of data grouped into samples of~100, 200, 500, and
     1000 showers, into samples of 24 hour duration, into samples of
     all continuous intervals of the array running (duration from~2
     to 21~days) and a sample formed for the total data set.  The
     analysis was performed in two ways: with the barometric effect
     taken into account, and without considering this effect (see
     Fomin et~al.\ (1999) for the details).  The homogeneity
     hypothesis was checked for confidence levels equal to~0.1, 0.05,
     0.02, and~0.01.

     It was found that among samples of~100 and 200 showers there
     exist several samples for which the homogeneity assumption
     cannot be accepted even for the smallest confidence level.  It
     is interesting to mention that we have found certain groups of
     days (e.g., in May~1998) such that the number of non-homogeneous
     samples grows at the beginning of the period and then decreases.
     We did not manage to reveal any methodical reasons of this
     behavior.  Moreover, we have not found any correlation between
     non-homogeneous data samples and EAS clusters; no correlation
     has also been revealed between non-homogeneous samples and
     sequences of showers with large delays (more that 1.5~min) (see
     Fomin et~al., (2001a)).  This allows us to conjecture that this
     effect reflects some astrophysical phenomena.

     Next, it was found out that the homogeneity assumption cannot be
     accepted (with the confidence level~0.01) for certain samples
     of~3 and more days of regular observation and also for the
     sample constructed of total data.  In particular, the series
     number for the total data set is equal to 829841, while the
     upper and low boundaries of series number that allow one to
     accept the homogeneity hypothesis with the given confidence
     level are 832580 and 835910 respectively.  But if we study the
     same data sample taking into account the barometric effect, then
     we may accept the homogeneity assumption with the confidence
     level equal to~0.05.

     For homogeneous samples, we have studied whether intervals
     between successive arrival times obey the exponential
     distribution.  This was performed by means of the
     $\chi^2$--test.  We have studied separately time intervals with
     lengths $\le30,$ 40, 50, 60, 90, and 120~s.  Data samples were
     grouped into bins of lengths~1, 2, 5, and 10~s.  It was found
     that for all homogeneous data sets there exists a time scale (as
     a rule, $\le90$~s) such that the hypothesis under consideration
     may be accepted at least at the confidence level~0.005.  In
     particular, we have analysed a set of EAS obtained during~51
     days of the array running for which the average atmospheric
     pressure was within 740--745~mm~Hg, and daily variation did not
     exceed 10~mm.  It was obtained that this sample (416998 showers)
     agrees with the homogeneity assumption at the confidence
     level~0.05, and the distribution of time intervals that are not
     longer than 55~s is exponential with the confidence level~0.9
     (see Fig.~1).

\begin{figure}[t]
 \includegraphics[width=8.5cm]{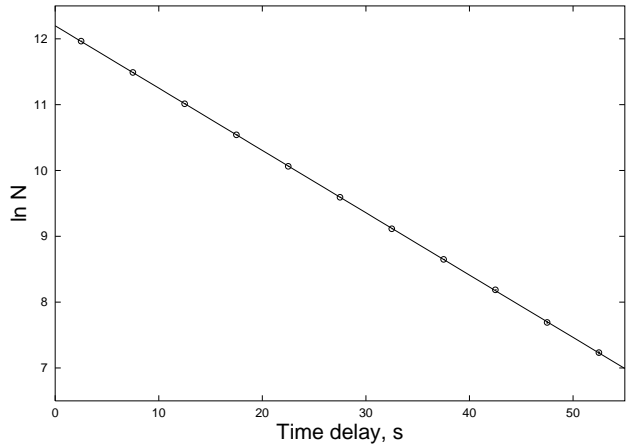}
 \caption{Distribution of time delays between 416998 showers (see the text);
  $N$ is the number of delays within a 5-seconds bin.}
\end{figure}

     In addition, we have studied EAS arrival times by methods of
     cluster analysis.  This investigation has revealed that there
     are definite sequences of EAS that can be identified as
     clusters.  The results will be reported in details elsewhere
     (Fomin et~al., 2001a).  We have also applied certain methods of
     nonlinear time series analysis to the data sets that contain
     clusters.  It was found out that the time series under
     consideration demonstrate chaotic dynamics (Fomin et~al.,
     2001b).  Finally, we have begun to study a more complete set of
     experimental data, which covers the period up to November~1999
     and contains $3.5\times10^6$ EAS.

%----------------------------------------------------------------------

\section{Conclusion}
     It should be noted that our results do not contradict the
     conclusions of other research groups (Chikawa et~al., 1991;
     Tsuji et~al., 1993).  However, it is necessary to increase
     shower statistics and to use other non-parametric methods of
     hypothesis verification since different criteria may give
     different results for the homogeneity assumption test (Bendat
     and Piersol, 1986).

%----------------------------------------------------------------------

\begin{acknowledgments}
     This work was done with financial support of The Federal
     Scientific-Technical Program for 2001, ``Research and design in
     the most important directions of science and techniques for
     civil applications,'' subprogram ``High Energy Physics,'' and
     grant of RFBR 99--02--16250.
\end{acknowledgments}

%----------------------------------------------------------------------

\end{document}